\documentclass[aps,prl,twocolumn,amsmath,amssymb,showpacs,superscriptaddress,notitlepage,longbibliography,10pt]{revtex4-1}

\usepackage[colorlinks=true,linkcolor=blue,anchorcolor=red,citecolor=blue, urlcolor=blue]{hyperref}
\usepackage{bm}
\usepackage{graphicx}
\usepackage{color}
\usepackage{multirow}
\usepackage{extarrows}
\usepackage{pifont}
\usepackage{cases}
\usepackage{cancel}
\usepackage[normalem]{ulem}

\begin{document}

\title{Parity-Selective Quantum Interference in Unconventional Magnets
}

\author{Yiyuan Chen}
\email{Equal contribution.}
\affiliation{Quantum Science Center of Guangdong-Hong Kong-Macao Greater Bay Area (Guangdong), Shenzhen 518045, China}

\author{Xiao-Bin Qiang}
\email{Equal contribution.}
\affiliation{State Key Laboratory of Quantum Functional Materials, Department of Physics, Guangdong Provincial Key Laboratory of Topological Matter, and Guangdong Basic Research Center of Excellence for Quantum Science, Southern University of Science and Technology (SUSTech), Shenzhen 518055, China}
\affiliation{Quantum Science Center of Guangdong-Hong Kong-Macao Greater Bay Area (Guangdong), Shenzhen 518045, China}

\author{Hai-Zhou Lu}
\email{Corresponding author: luhz@sustech.edu.cn}
\affiliation{State Key Laboratory of Quantum Functional Materials, Department of Physics, Guangdong Provincial Key Laboratory of Topological Matter, and Guangdong Basic Research Center of Excellence for Quantum Science, Southern University of Science and Technology (SUSTech), Shenzhen 518055, China}
\affiliation{Quantum Science Center of Guangdong-Hong Kong-Macao Greater Bay Area (Guangdong), Shenzhen 518045, China}

\author{Congjun Wu}
\email{Corresponding author: wucongjun@westlake.edu.cn}
\affiliation{New Cornerstone Science Laboratory, Department of Physics, Institute for Theoretical Sciences, and Key Laboratory for Quantum Materials of Zhejiang Province, Westlake University, Hangzhou 310024, China}


\author{X. C. Xie}
\affiliation{International Center for Quantum Materials, School of Physics, Peking University, Beijing 100871, China}
\affiliation{Institute for Nanoelectronic Devices and Quantum Computing, Fudan University, Shanghai 200433, China}
\affiliation{Hefei National Laboratory, Hefei 230088, China}

\begin{abstract}
Nonrelativistic spin splitting has emerged as a defining feature of unconventional magnetic states.  
Here we show that quantum interference offers a symmetry-resolved probe for the parity of such spin splitting, as even- and odd-parity magnets exhibit qualitatively different weak (anti-)localization behaviors. 
Even-parity magnets can host time-reversal symmetric scattering loops depicted by the spin group, giving rise to a weak localization that can be suppressed by spin-orbit coupling.
In contrast, odd-parity magnets have a hidden weak antilocalization that can be released by spin-orbit coupling. 
We further derive the corresponding magnetoconductivity formulas to facilitate experimental explorations.
Our results not only advance the weak (anti-)localization theory to unconventional magnets but also suggest a practical diagnostic of unconventional spin-split magnetic states.
\end{abstract}

\maketitle

{\color{blue}\emph{Introduction}} - Non-relativistic spin splitting \cite{wu2004,wu2007,Hayami19jpsj,LiuJW21NC} magnet depicted by 
spin space groups \cite{Liu22prx,Libor22PRX2,Xiao24prx} has emerged as a paradigm in magnetism. 
Recent studies have identified a diverse library of candidate materials, classified primarily according to the 
parity of spin splitting in momentum space: even-parity altermagnets  (e.g., $d$-wave, $g$-wave, ...)~\cite{Libor22PRXLandscape,Naka19NC,Ahn19PRB,yuanLD20PRB,Libor20SciAdv,ShaoDF21NC,Bai24AFM,Osumi24PRB,Krempask24Nat,Lee24PRL,Reimers24NC,Ding24prl,Zhang25np,Hayami20prb2} and odd-parity magnets (e.g., $p$-wave, $f$-wave, ...)~\cite{Hayami20prb1,Hayami20prb2,Martin12prb,Hayami22prb,Kudasov24prb,Anna23arxiv,Brekke24prl,Chakraborty25nc,Song25nature,Yamada25nature}. 
Nevertheless, compelling experimental signature of these unconventional magnets is still a topic of tremendous interest \cite{Libor22PRX1,Rafael21PRL,Bose22NE,Pari25prb,Ezawa25prb,Hiraishi24PRL,Kessler24npj,Liu24PRL,ZiHL24arxiv,Lin25prl}. 
Symmetry can always give decisive criteria for states of condensed matter. Weak localization (WL) and weak antilocalization (WAL) are highly sensitive to symmetry, appearing as low-temperature 
positive and negative 
magnetoconductivity, respectively,
in metals \cite{Bergman82prl}, semiconductors \cite{Rosenbaum81prl1,Rosenbaum81prl2,Bishop82prb,Miller03prl}, 2D materials  \cite{Wu07prl,Tikhonenko09prl,Yuan13np,Neal13ACS}, and topological materials \cite{Checkelsky09prl,Peng10natmat,Chen10prl,He11prl,Wang11prb,Liu11prb,Liu12prl,Kim13prl,HuangXC15prx,Li16np,ZhangCL16nc}. 
However, weak (anti-)localization arise from the quantum interference between time-reversal scattering loops, which are highly nontrivial 
to be defined in magnets \cite{Neumaier07prl,Dugaev08prl}.


\begin{figure}[tp]
\includegraphics[width=0.9\linewidth]{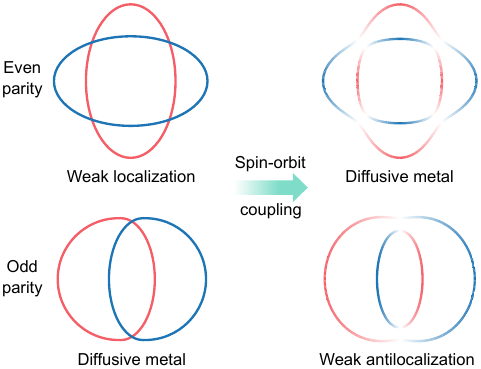}
\caption{The spin-up (red) and spin-down (blue) Fermi surfaces of even-parity altermagnets and odd-parity $p$-waves magnets in the absence and presence of spin-orbit coupling. In absence of spin-orbit coupling, altermagnets may exhibit weak localization (WL) and $p$-wave magnets are diffusive metals. Spin-orbit coupling can suppress WL in altermagnets and release weak antilocalization (WAL) in $p$-wave magnets.}
\label{Fig:Sym}
\end{figure}

In this Letter, we investigate the weak (anti-)localization in even- and odd-parity magnets, focusing on 
the definition of time-reversal symmetry in the context of spin groups and influence of spin-orbit coupling. 
In the absence of spin-orbit coupling, we figure out the spin-group symmetries that support the time-reversal scattering loops and WL and WAL channels in even- and odd-parity magnets. 
Specifically, even-parity $d$-wave altermagnets can host two WL channels that add up to each other, 
while odd-parity $p$-wave magnets have a WL channel and a WAL channel that exactly cancel each other. 
Hence, we predict WL in altermagnets and no WL or WAL in $p$-wave magnets in the absence of spin-orbit coupling. Spin-orbit coupling can suppress the WL channels in both $d$-wave altermagnets and $p$-wave magnets, leading to the suppression of WL in $d$-wave altermagnets and activation of WAL in $p$-wave magnets in the presence of spin-orbit coupling, as summarized in Figure~\ref{Fig:Sym} and
Table~\ref{Tab:symmetry}.
More importantly, our conclusions also apply to higher-order even- and odd-parity magnets (e.g., $f$-, $g$- or $i$-wave magnets). Our results  advance the weak (anti-)localization theory and will be helpful for explorations on unconventional magnets.

\begin{table}[bthp]
\caption{Channels of weak localization (WL) and weak antilocalization (WAL) of even- and odd-parity magnets and their spin-group symmetries in the absence and presence of spin-orbit coupling (SOC). We adopt the notation of spin group $\mathcal{T}[A||B]$, $\mathcal{T}$ is the overall time-reversal operation, $A$ is for spin degree of freedom, $B$ is for real space, $C_{2x(y)}$ is two-fold spin rotation around the $x$ ($y$) axis, and $E$ is identity. The $+$ and $-$ signs mean that the channels add up or cancel each other.}
\label{Tab:symmetry}
\centering
\setlength{\tabcolsep}{0.15cm}
\renewcommand\arraystretch{1.2}
\begin{tabular}{cccccccccccccc}
\hline
\hline
&&without SOC&with SOC
\\
\hline
 &$\mathcal{T}[C_{2x}||E]$ (WL) &\ding{51}&\ding{55} \\
Even &   $+$   & 
\\
 &$\mathcal{T}[C_{2y}||E]$ (WL)  &\ding{51}&\ding{55}
\\
\hline
&$\mathcal{T}[C_{2z}||E]$ (WL)   &\ding{51}& \ding{55}
\\ 
Odd & $-$ &\\
&$\mathcal{T}$ (WAL)   &\ding{51}&\ding{51}
\\
\hline
\hline
\end{tabular}
\end{table}

{\color{blue}\emph{Magnets with even- and odd-parity spin splitting}} - The Hamiltonian that describes electrons in 2D magnets with even- and odd-parity non-relativistic spin splitting can be given as
\begin{eqnarray}\label{Eq: Hamiltonian}
H(\mathbf{k})=a k^2 +m_\mathbf{k}
\sigma_z+\lambda(k_x\sigma_y-k_y\sigma_x),
\end{eqnarray}
where the $a$ term represents the spin-independent parabolic band and $k^2=k_x^2+k_y^2$, $\lambda$ measures the spin-orbit coupling, and $m_\mathbf{k}$ describes the non-relativisitic spin spitting with different orders of symmetries ($d$, $g$, $i$ for even-parity and $p$, $f$ for odd-parity) \cite{Ezawa26prb}
\begin{eqnarray}
    m^p_\mathbf{k}&=&Jk_x,\nonumber\\
    m^d_\mathbf{k}&=&J(k_x^2-k_y^2),\nonumber\\
    m^f_\mathbf{k}&=&Jk_x(k_x^2-3k_y^2),\nonumber \\
    m^g_\mathbf{k}&=&Jk_xk_y(k_x^2-k_y^2), \nonumber  \\
    m^i_\mathbf{k}&=&Jk_xk_y(k_x^2-3k_y^2)(k_y^2-3k_x^2). \label{eq:UM}
\end{eqnarray}
For instance, altermagnets are of even parity and $p$-wave magnets are of odd parity. Because unconventional magnets with non-relativistic spin splitting are antiferromagnets with zero net magnetization, the integral of $m_\mathbf{k}$ on any isoenergy contour is supposed to be zero.

{\color{blue}\emph{Time-reversal symmetry in context of spin group}} - WL or WAL arises from quantum interference between time-reversed loops formed by trajectories of electrons as they are scattered by disorder in solids \cite{Bergmann84pr,Lee85rmp,Akkermans2007}, so time-reversal symmetry is required. However, it is always nontrivial to define time-reversal symmetry in magnets. We will show that, with the help of spin group, we can introduce the anti-unitary operators 
to depict the effective time-reversal symmetries for the unconventional magnets with non-relativistic spin splitting.

For $m_\mathbf{k}$ of even parity  (e.g., altermagnets), if in the absence of spin-orbit coupling ($\lambda=0$), there are two anti-unitary operators 
\begin{eqnarray}
    \mathcal{T}_1=\sigma_zK, \ \ \ \ \mathcal{T}_2=K,
\end{eqnarray}
with $K$ the
complex conjugate, so that
$\mathcal{T}_1H(\mathbf{k})\mathcal{T}_1^{-1}=H(-\mathbf{k})$ and $\mathcal{T}_2H(\mathbf{k})\mathcal{T}_2^{-1}=H(-\mathbf{k})$. They depict the effective time-reversal symmetries 
\begin{eqnarray}
\mathcal{T}[C_{2x}||E],\ \ \ \ \mathcal{T}[C_{2y}||E],    
\end{eqnarray}
respectively, of altermagnets \cite{Libor22PRXLandscape} in the absence of spin-orbit coupling, where 
$\mathcal{T}$ is the overall time-reversal operation and $[C_{2x/2y}||E]$ is the spin group operation that rotates spin by $\pi$ around the $x/y$ axis and remains identity in real space. Because $\mathcal{T}_1^2=1$ and $\mathcal{T}_2^2=1$, they belong to the orthogonal class according to the symmetry classification of disordered ensembles~\cite{Dyson62jmp}. The quantum interference of the orthogonal class is WL \cite{Hikami80ptp}. Spin-orbit coupling forbids the free spin rotation so it breaks $\mathcal{T}_{1}$ and $\mathcal{T}_{2}$, leading to the suppression of WL.

For $m_\mathbf{k}$ of odd parity (e.g., $p$-wave magnets), if in the absence of spin-orbit coupling ($\lambda=0$), there are two anti-unitary operators 
\begin{eqnarray}
    \mathcal{T}_3=\sigma_x K, \ \ \ \ \mathcal{T}_4=\sigma_y K,
\end{eqnarray}
so that $\mathcal{T}_3H(\mathbf{k})\mathcal{T}_3^{-1}$ $=$ $H(-\mathbf{k})$ and $\mathcal{T}_4H(\mathbf{k})\mathcal{T}_4^{-1}=H(-\mathbf{k})$.
They depict the effective time-reversal symmetries 
\begin{eqnarray}
    \mathcal{T}[C_{2z}||E],\ \ \ \ \mathcal{T}
\end{eqnarray}
respectively, in the absence of spin-orbit coupling. Because $\mathcal{T}_3^2=1$, it also belongs to the orthogonal class that gives rise to WL. In contrast, $\mathcal{T}_4^2=-1$ and it belongs to the symplectic class~\cite{Dyson62jmp}. 
The quantum interference of the symplectic class is WAL \cite{Hikami80ptp}.
The weak localization from $\mathcal{T}_3$ cancels with weak antilocalization from $\mathcal{T}_4$, leading to vanishing WL or WAL in the absence of spin-orbit coupling \cite{Supp}.
Spin-orbit coupling forbids the free spin rotation
so it breaks $\mathcal{T}[C_{2z}||E]$ but preserves $\mathcal{T}$, therefore, WL is suppressed while WAL is released by spin-orbit coupling.

{\color{blue}\emph{Feynman diagrams and Cooperon gaps}} - Now we use diagrammatic calculations \cite{Bergmann84pr, Lee85rmp} to verify the above physical picture. Although the calculation is routine, the application to the magnets with non-relativistic spin splitting is nontrivial.
Different from the Drude conductivity arising from the particle-hole correlation, the quantum interference is from particle-particle correlation, similar to the Cooper pairing of electrons, so the quantum interference is also referred to as the diffusion of Cooperons. The electron has a spin of 1/2, so as two paired electrons, a Cooperon can have the total angular momenta of 1 and 0, i.e., the triplet and singlet Cooperons \cite{Altshuler81jetp}. Each of the triplet Cooperons gives a WL channel and the singlet Cooperon gives one WAL channel. Because the triplet Cooperons have a finite angular momentum of 1, they can be suppressed by the spin-orbit coupling which behaves like a spin-dependent ``magnetic field". The singlet Cooperon therefore is robust against spin-orbit coupling because it has no angular momentum.



\begin{figure}[bp]
\includegraphics[width=0.48\textwidth]{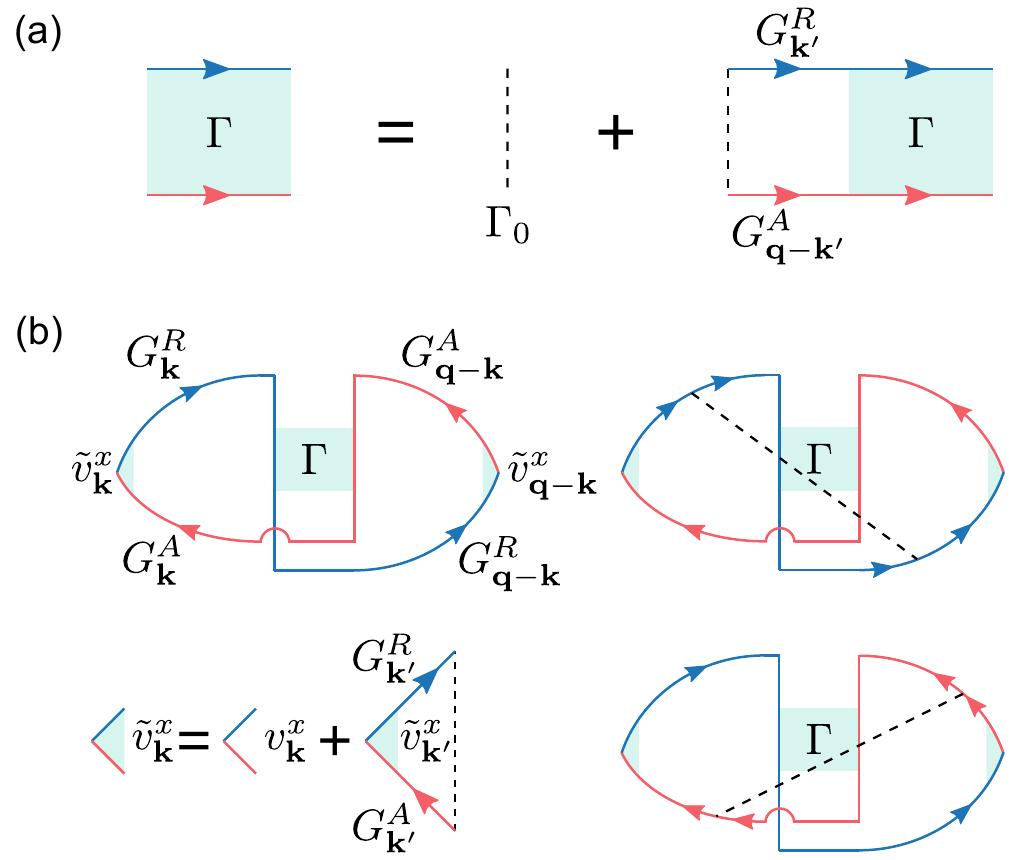}
\caption{The diagrams for the quantum interference correction to conductivity \cite{Hikami80ptp,Supp,Altshuler81jetp,McCann06prl,Fu19prl,
Wang20prl,Shon98jpsj}. The arrowed
solid lines and dashed lines represent the Green’s functions $G^{R/A}$ and impurity scattering, respectively. (a) The Bethe-Salpeter equation for Cooperon $\Gamma$ of maximally crossed diagrams.
(b) The bare and two dressed maximally crossed diagrams give the quantum conductivity correction. $\tilde{v}$ is the vertex correction of the velocity $v$ by ladder diagrams.}
\label{Fig:Feymann diagram}
\end{figure}

In the language of diagrammatic techniques, the suppression of a WL or WAL channel is described by a finite Cooperon gap. A vanishing Cooperon gap leads to a divergent contribution to the conductivity to support a WL or WAL channel. The Cooperons are calculated with the help of the diagrams of Bethe-Salpeter equation in Fig.~\ref{Fig:Feymann diagram}(a), where the arrow solid lines are 
the disorder-averaged retarded/advanced Green's functions defined in the $\sigma_z$-eigenbasis \cite{McCann06prl,Fu19prl,Wang20prl} as
$  G_{\mathbf{k}}^{R/A}=(g^{R/A}_{\mathbf{k}})^{-1}[E_F-ak^2+m_\mathbf{k}\sigma_z+\lambda (k_x\sigma_y-k_y\sigma_x)]$ and $g^{R/A}_{\mathbf{k}}=(E_F\pm i\Delta-ak^2)^2-m_\mathbf{k}^2-\lambda^2k^2$, $E_F$ is the Fermi energy, $\Delta=\Gamma_0\rho\pi/2$ is the disorder averaged self-energy with
$\Gamma_0$ the bare Cooperon and $\rho$ the density of states at the Fermi energy. 
The Bethe-Salpeter equation of the full Cooperon $\Gamma(\mathbf{q})$ reads 
\begin{eqnarray}
\Gamma(\mathbf{q})=\left[\Gamma_0^{-1} -\Pi(\mathbf{q})\right]^{-1},
\end{eqnarray}
where the Kernel function $\Pi(\mathbf{q})=\sum_{\mathbf{k}}G^R_{\mathbf{k}}\otimes G^A_{\mathbf{q}-\mathbf{k}}$. The gaps of Cooperon in different channels can be evaluated from $[\Gamma(\mathbf{q})]^{-1}$ as $\mathbf{q}\rightarrow 0$. Above we have used the particle-particle basis $\{|{\uparrow}{\uparrow}\rangle,|{\uparrow\downarrow}\rangle,|{\downarrow\uparrow}\rangle,|{\downarrow\downarrow}\rangle\}$.
More conveniently, they can be transformed into the singlet and triplet basis $\{|1,1\rangle,|1,0\rangle,|1,-1\rangle, |0,0\rangle \}$ via a transformation $\mathcal{U}$ \cite{Supp}.

For even-parity magnets, e.g.,
$m_\mathbf{k}^d$ in Eq. (\ref{eq:UM}), 
we arrive at
$\mathcal{U}^\dagger[\Gamma(0)]^{-1}\mathcal{U}=
\sum_\mathbf{k}
(g^{R}_{\mathbf{k}} g^{A}_{\mathbf{k}})^{-1} M(\mathbf{k})$, where because $(g^{R}_{\mathbf{k}} g^{A}_{\mathbf{k}})^{-1}$ is diagonal, the Cooperon gaps is determined by the matrix $M(\mathbf{k})$, which reads
\begin{eqnarray}\label{eq:M_d}
   \begin{bmatrix}
    \lambda^2k^2&0&-\lambda^2 k_-^2 &0\\
    0&2m_{\mathbf{k}}^2+2\lambda^2k^2&0&0\\
    -\lambda^2 k_+^2 &0& \lambda^2k^2 &0\\
    0&0&0& 2m_{\mathbf{k}}^2 \\
    \end{bmatrix}.
\end{eqnarray}
When spin-orbit coupling $\lambda$ is zero, $M(\mathbf{k})$ becomes diagonal and the $\lambda^2k^2$ terms vanish, leading to closing of the Cooperon
gaps of two triplet channels $|1,1\rangle$ and $|1,-1\rangle$. Hence, WL dominates in even-parity magnets when there is no spin-orbit coupling. 
When spin-orbit coupling is turned on, it can be shown that the matrix $\mathcal{U}^\dagger[\Gamma(0)]^{-1}\mathcal{U}$ has only positive eigen values, which means that all four channels have finite Cooperon gaps so there is no WL or WAL. 

For odd-parity magnets, e.g., $m_\mathbf{k}^p$ in Eq. (\ref{eq:UM}), the matrix 
$M(\mathbf{k})$ that determines the Cooperon gaps reads
\begin{eqnarray}
    \begin{bmatrix}
    2m_{\mathbf{k}}^2+\lambda^2k^2&i\sqrt2\lambda m_{\mathbf{k}}k_-& -\lambda^2
    k_-^2&0\\
    -i\sqrt2\lambda m_{\mathbf{k}}k_+&2\lambda^2k^2&-i\sqrt2\lambda m_{\mathbf{k}}k_-&0\\
    -\lambda^2 k_+^2&i\sqrt2\lambda m_{\mathbf{k}}k_+&2m_{\mathbf{k}}^2+\lambda^2k^2&0\\
    0&0&0&0&
    \end{bmatrix},
\end{eqnarray}
where the 0 matrix elements  mean that the Cooperon gap of the singlet channel $|0,0\rangle$ does not couple to the triplet channels and always has a  vanishing Cooperon gap to contribute a WAL channel.
At zero spin-orbit coupling $\lambda$, the gap of the triplet channel $|1,0\rangle$ also vanishes, giving rise to a WL channel that cancels
the WAL of the singlet channel. Hence, there is no WL or WAL when there is no spin-orbit coupling \cite{Supp}. At finite spin-orbit coupling $\lambda$, it can be proved that the upper $3\times 3$ block of $\mathcal{U}^\dagger[\Gamma(0)]^{-1}\mathcal{U}$ has only positive eigen values, hence, all the triplet Cooperon gaps are finite, leading to the suppression of WL. 
As a result, WAL of the singlet channel dominates in the odd-parity magnets in the presence of spin-orbit coupling.

{\color{blue}\emph{Magnetoconductivity and experimental implications}} 
- 
Below we show that WL of even-parity  $d$-wave altermagnets and WAL of odd-parity $p$-wave magnets can be measured by the low-field magnetoconductivity. 
 
\begin{figure}[tp]
\includegraphics[width=0.49\textwidth]{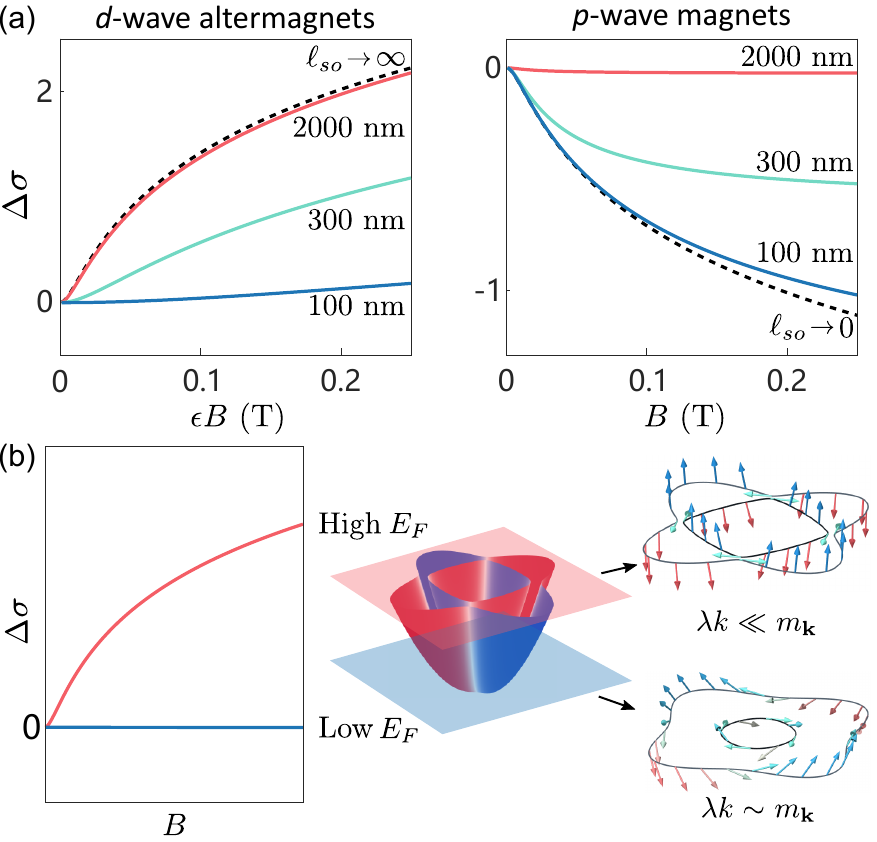}
\caption{(a) Magnetoconductivity $\Delta \sigma$ of the $d$-wave altermagnets [in units of $(1+\epsilon^2)e^2/(4\epsilon \pi^2 \hbar)$] and $p$-wave magnets [in units of $e^2/(2 \pi^2 \hbar)$] at different strengths of spin-orbit scattering. A shorter spin-orbit relaxation length $\ell_{so}$ means stronger spin-orbit scattering. Spin-orbit scattering suppresses WL in the $d$-wave altermagnets and activates WAL in the $p$-wave case. 
Parameters are chosen as $\ell=10$ nm and $\ell_\phi=400$ nm.
(b) For $d$-wave altermagnets, the Fermi surface is dominated by $m_{\mathbf{k}}$ for high Fermi energy, corresponding to WL. For low Fermi energy, $m_{\mathbf{k}}$ become relatively close to the spin-orbit coupling. WL is suppressed and the system is in a diffusive metal state.}
\label{Fig:Magnetoconductance}
\end{figure}

For $d$-wave altermagnets characterized by $m^d_\mathbf{k}$
in Eq. (\ref{eq:UM}), 
the quantum interference correction to conductivity can be calculated with the help of the diagrams in Fig.~\ref{Fig:Feymann diagram} (b) as
\begin{eqnarray}\label{Eq:Dwave}
    &&\delta\sigma^d_{xx} =-\frac{2e^2\ell^2}{h}\int  \frac{d^2\mathbf{q}}{(2\pi)^2}
   \nonumber \\
    &\times& 
\Big[\frac{1}{\frac{4\ell^2}{3\ell_{so}^2}+\ell^2(q_x^2+\epsilon^2q_y^2)}+\frac{\epsilon^2}{\frac{4\ell^2}{3\ell_{so}^2}+\ell^2(\epsilon^2q_x^2+q_y^2)}\Big],
    \nonumber \\
\end{eqnarray}
where $\ell=\sqrt{D_d\tau}$, $\ell_{so}=\sqrt{D_d\tau_{so}}$, $\ell_\phi=\sqrt{D_d\tau_\phi}$ are the mean free path, spin-orbit relaxation length, and phase coherence length, respectively, the diffusion constant $D_d=(v^d_F)^2\tau/2$,
the Fermi velocity $v^d_F=\sqrt{4(a+J)E_F/\hbar^2}$,
and $\epsilon^2=(a+J)/(a-J)$,  $a$ and $J$ depict the parabolic band dispersion defined in Eq. (\ref{Eq: Hamiltonian}) and
anisotropy induced by exchange interaction defined in Eq. (\ref{eq:UM}), respectively. For convenience, we describe the spin-orbit coupling in terms of impurity scattering, i.e., spin-orbit scattering. The two terms in Eq. (\ref{Eq:Dwave}) stand for the two WL channels that can be gapped out by the spin-orbit scattering if the Cooperon gaps $4\ell^2/3\ell_{so}^2$ are finite, consistent with the above symmetry analysis and Cooperon gap calculations. By physically considering the lower and upper bounds as $1/\ell_\phi$ and $1/\ell$
in Eq. (\ref{Eq:Dwave}), we can find the quantum interference correction to conductivity at zero magnetic field as
$\delta\sigma^d_{xx}(0)\simeq
-\alpha e^2/(2\pi^2\hbar)
\ln (1+\ell_2^2/\ell^2)$
and at finite $z$-direction magnetic field $B$ as 
$\delta\sigma^d_{xx}(B)
\simeq
\alpha e^2/(2\pi^2\hbar)[\Psi (\ell_B^2/\epsilon \ell_2^2+1/2) - \ln(\ell_B^2/\epsilon \ell_1^2)]$,
where  $\alpha\equiv (1+\epsilon^2)/2\epsilon$, the magnetic length $\ell_B =\sqrt{\hbar/(4eB)}$, 
$1/\ell_1^2 \equiv 
1/\ell^2+4/3\ell_{so}^2+
1/\ell_\phi^2$, $1/\ell_2^2\equiv 
4/3\ell_{so}^2+1/\ell_\phi^2$.
$\Psi(x)$ is the digamma function $
\Psi(x)=-\gamma-1/x
-\sum_{k=1}^\infty 
[1/(x+k)-1/k]$ with the Euler constant $\gamma\approx 0.57721$ \cite{Arfken2012,Gradshteyn}. 
In the limit of vanishing spin-orbit scattering, i.e., $ 1/\ell_{so} \rightarrow 0 $,
Eq. (\ref{Eq:Dwave}) describes the WL at $B=0$, which can be measured by the magnetoconductivity $\Delta\sigma^d_{xx}(B)\equiv \delta\sigma^d_{xx}(B)-\delta\sigma^d_{xx}(0)$ and
\begin{eqnarray}
   \Delta\sigma^d_{xx}(B) \simeq \alpha\frac{e^2}{2\pi^2\hbar}
\Big[\Psi\Big(\frac{\ell_B^2}{\epsilon\ell_\phi^2}+\frac{1}{2}\Big)-\ln \frac{\ell_B^2}{\epsilon\ell_\phi^2}\Big], 
\end{eqnarray}
In contrast, if spin-orbit scattering is strong,  i.e., $\ell_\phi, \ell_B \gg  \ell_{so}\sim \ell$, the system turns to a diffusive metal at $B=0$, then 
the magnetoconductivity nearly vanishes, i.e., $ \Delta\sigma^d_{xx}(B)\rightarrow 0$, as shown in Fig. \ref{Fig:Magnetoconductance}(a).
  

For $p$-wave magnets characterized by $m^p_\mathbf{k}$ in Eq. (\ref{eq:UM}), the quantum interference correction to conductivity can be found as
\begin{eqnarray}\label{Eq:Pwave}
\delta\sigma^p_{xx} =-\frac{2e^2\ell^2}{h}\int\frac{d^2\mathbf{q}}{(2\pi)^2}\Big(\frac{1}{\frac{4\ell^2}{3\ell^2_{so}}+\ell^2 q^2}-\frac{1}{\ell^2 q^2}\Big),    
\end{eqnarray}
where $\ell=\sqrt{D_p\tau}$, $\ell_{so}=\sqrt{D_p\tau_{so}}$, $\ell_\phi=\sqrt{D_p\tau_\phi}$ are the mean free path, spin-orbit relaxation length, and phase coherence length, respectively, the diffusion constant $D_p=(v^p_F)^2\tau/2$,
the Fermi velocity $v^p_F=\sqrt{4aE_F/\hbar^2}$. The first and second terms in Eq. (\ref{Eq:Pwave}) describe a WL channel that can be gapped out by the spin-orbit scattering and a WAL channel, respectively. The quantum interference correction to conductivity at zero magnetic field is found as $\delta\sigma^p_{xx}(0) \simeq  
e^2/(4\pi^2\hbar) 
[
\ln ( \ell_\phi^2/\ell^2)- 
\ln(1+\ell_2^2/\ell^2)]$, where the first and second terms represent the WL and WAL contributions, respectively. The quantum interference corrections to conductivity at finite $z$-direction magnetic field is found as
$\delta\sigma^p_{xx}(B)
    \simeq  e^2/(4\pi^2\hbar)
     \{\Psi (\ell_B^2/\ell_2^2+1/2)
    -\ln(\ell_B^2/\ell_1^2)-[\Psi(\ell_B^2/\ell_{\phi}^2+1/2)-\ln(\ell_B^2/\ell^2)]
     \}$.
In the case of strong spin-orbit scattering, i.e., $\ell_\phi, \ell_B \gg  \ell_{so}\sim \ell$, the WL channel is gapped out, only the WAL channel survives, which can be measured by the magnetoconductivity 
$\Delta\sigma^p_{xx}(B)\equiv \delta\sigma^p_{xx}(B)-\delta\sigma^p_{xx}(0)$ and
\begin{eqnarray}
   \Delta\sigma^p_{xx}(B)\approx \alpha\frac{e^2}{2\pi^2\hbar}\Big[\Psi\Big(\frac{\ell_B^2}{\ell_\phi^2}+\frac{1}{2}\Big)-\ln\frac{\ell_B^2}{\ell_\phi^2}\Big], \ \ \
\end{eqnarray}
with $\alpha=-1/2$.
By comparison, in the limit of vanishing spin-orbit scattering, i.e., $ 1/\ell_{so} \rightarrow 0 $, the magnetoconductivity vanishes, i.e., $\Delta\sigma^p_{xx}(B)\rightarrow 0$, as shown in Fig. \ref{Fig:Magnetoconductance}(a).

{\color{blue}\emph{Discussions}} - Besides directly tuning the spin-orbit coupling strength $\lambda$, the relative magnitude of $m_{\mathbf{k}}$ and $\lambda k$ can also be controlled by shifting the Fermi surface. For a higher-order magnetic splitting $m_{\mathbf{k}}\propto k^n$ ($n>1$), lowering the Fermi energy toward the band bottom reduces $m_{\mathbf{k}}$ faster than $\lambda k$. As illustrated in Fig. \ref{Fig:Magnetoconductance}(b), the system can therefore evolve from a regime with $\lambda k\ll m_{\mathbf{k}}$, where WL is pronounced, to one with $\lambda k\sim m_{\mathbf{k}}$, where the WL Cooperon gaps are substantially enhanced and the system approaches a diffusive-metal regime. Thus, tuning Fermi energy provides an alternative route to control quantum interference.

For $p$-wave magnets, two triplet Cooperon channels $|1,1\rangle$ and $|1,-1\rangle$ may also give rise to WL at finite wavevectors $\pm\mathbf{q}_0=\pm J/(2a)\hat x$,  
because the spin-up and spin-down Fermi surfaces of $p$-wave magnets are shifted in opposite directions. In this case, $p$-wave magnets may have totally 3 triplet Cooperon channels and 1 singlet Cooperon channel in the absence of spin-orbit coupling to exhibit WL. Spin-orbit coupling can suppress the 3 WL channels, leading to a crossover to WAL. 
Nevertheless, in realistic lattices, the warping of the Fermi surface may suppress the $|1,1\rangle$ and $|1,-1\rangle$ WL channels by breaking the $\mathbf{k}\rightarrow \pm \mathbf{q}_0-\mathbf{k}$ symmetry.



\bibliographystyle{apsrev4-1-etal-title_10authors}
\bibliography{refs-RuO2,refs-transport}

\end{document}